\documentclass[12pt]{article}
\usepackage{graphicx}

\def\hybrid{\topmargin 0pt      \oddsidemargin 0pt
        \headheight 0pt \headsep 0pt
       \voffset-1cm
        \textwidth 6.25in       
       \textheight 9.5in       
        \marginparwidth 0.0in
        \parskip 5pt plus 1pt   \jot = 1.5ex}
\catcode`\@=11
\def\marginnote#1{}

\newcount\hour
\newcount\minute
\newtoks\amorpm
\hour=\time\divide\hour by60
\minute=\time{\multiply\hour by60 \global\advance\minute by-\hour}
\edef\standardtime{{\ifnum\hour<12 \global\amorpm={am}%
        \else\global\amorpm={pm}\advance\hour by-12 \fi
        \ifnum\hour=0 \hour=12 \fi
        \number\hour:\ifnum\minute<10 0\fi\number\minute\the\amorpm}}
\edef\militarytime{\number\hour:\ifnum\minute<10 0\fi\number\minute}

\def\draftlabel#1{{\@bsphack\if@filesw {\let\thepage\relax
   \xdef\@gtempa{\write\@auxout{\string
      \newlabel{#1}{{\@currentlabel}{\thepage}}}}}\@gtempa
   \if@nobreak \ifvmode\nobreak\fi\fi\fi\@esphack}
        \gdef\@eqnlabel{#1}}
\def\@eqnlabel{}
\def\@vacuum{}
\def\draftmarginnote#1{\marginpar{\raggedright\scriptsize\tt#1}}

\def\draftlabel#1{{\@bsphack\if@filesw {\let\thepage\relax
   \xdef\@gtempa{\write\@auxout{\string
      \newlabel{#1}{{\@currentlabel}{\thepage}}}}}\@gtempa
   \if@nobreak \ifvmode\nobreak\fi\fi\fi\@esphack}
        \gdef\@eqnlabel{#1}}
\def\@eqnlabel{}
\def\@vacuum{}
\def\draftmarginnote#1{\marginpar{\raggedright\scriptsize\tt#1}}

\def\draft{\oddsidemargin -.5truein
        \def\@oddfoot{\sl preliminary draft \hfil
        \rm\thepage\hfil\sl\today\quad\militarytime}
        \let\@evenfoot\@oddfoot \overfullrule 3pt
        \let\label=\draftlabel
        \let\marginnote=\draftmarginnote
   \def\@eqnnum{(\theequation)\rlap{\kern\marginparsep\tt\@eqnlabel}%
\global\let\@eqnlabel\@vacuum}  }

\def\numberbysection{\@addtoreset{equation}{section}
        \def\theequation{\thesection.\arabic{equation}}}

\def\underline#1{\relax\ifmmode\@@underline#1\else
        $\@@underline{\hbox{#1}}$\relax\fi}

\def\titlepage{\@restonecolfalse\if@twocolumn\@restonecoltrue\onecolumn
     \else \newpage \fi \thispagestyle{empty}\c@page\z@
        \def\thefootnote{\fnsymbol{footnote}} }

\def\endtitlepage{\if@restonecol\twocolumn \else  \fi
        \def\thefootnote{\arabic{footnote}}
        \setcounter{footnote}{0}}  
\relax

\numberbysection
\hybrid

\newfont{\Bbb}{msbm10 scaled 1\@ptsize00}
\newfont{\Bbbb}{msbm7 scaled 1\@ptsize00}
\newcommand{\CC}{\mbox{\Bbb C}}

\newcommand{\DDD}{\raise-1pt\hbox{$\mbox{\Bbbb D}$}}

\newcommand{\UUU}{\raise-1pt\hbox{$\mbox{\Bbbb U}$}}

\newcommand{\ZZ}{\mbox{\Bbb Z}}
\newcommand{\z}{\raise-1pt\hbox{$\mbox{\Bbbb Z}$}}

\newcommand{\SSS}{\mbox{\Bbb S}}
\newcommand{\sss}{\raise-1pt\hbox{$\mbox{\Bbbb S}$}}

\def\beq{\begin{equation}}
\def\eeq{\end{equation}}
\def\p{\partial}

\newtheorem{theorem}{Theorem}[section]
\newtheorem{lemma}{Lemma}[section]
\newtheorem{lemma-definition}{Lemma-Definition}[section]
\newtheorem{corollary}{Corollary}[section]

\newtheorem{remark}{Remark}[section]

\def\res{\mathop{\hbox{res}}\limits}

\def\square{\hfill
{\vrule height6pt width6pt depth1pt} \break \vspace{.01cm}}

\begin{document}

\begin{titlepage}

\title{Toda lattice with constraint of type B}

\author{I. Krichever\thanks{
Columbia University, New York, USA;
e-mail: krichev@math.columbia.edu}
\and
A.~Zabrodin\thanks{
Skolkovo Institute of Science and Technology, 143026, Moscow, Russia and
National Research University Higher School of Economics,
20 Myasnitskaya Ulitsa,
Moscow 101000, Russia and
NRC KI KCTEP, Moscow, Russia;
e-mail: zabrodin@itep.ru}}

\date{October 2022}
\maketitle

\vspace{-7cm} \centerline{ \hfill ITEP-TH-24/22}\vspace{7cm}

\begin{abstract}

We introduce a new integrable hierarchy of nonlinear differential-difference equations which is a subhierarchy of the 2D Toda lattice defined by
imposing a constraint to the Lax operators of the latter. The 
2D Toda lattice with the constraint can be regarded as a discretization
of the BKP hierarchy. We construct its algebraic-geometrical solutions 
in terms of Riemann and Prym theta-functions.

\end{abstract}

\end{titlepage}

\vspace{5mm}

%

\tableofcontents

\vspace{5mm}

\section{Introduction}

The 2D Toda lattice hierarchy \cite{UT84}
plays a very important role in the theory of integrable
systems. The commuting flows of the hierarchy 
are parametrized by infinite sets of complex time
variables ${\bf t}=\{t_1, t_2, t_3, \ldots \}$ (``positive times'') and $\bar {\bf t}=\{\bar t_1,
\bar t_2, \bar t_3, \ldots \}$ (``negative times''), together with the ``zeroth time'' $n\in \ZZ$.
Equations of the hierarchy are differential in the times 
${\bf t}$, $\bar {\bf t}$ and
difference in $n$. They can be represented in the Lax form as evolution
equations for two Lax operators $L$, $\bar L$ which are pseudo-difference
operators, i.e., half-infinite sums of integer powers of the shift
operators $e^{\pm \p_n}$ with coefficients depending on $n$ and ${\bf t}$,
$\bar {\bf t}$.
A common solution is provided by the tau-function 
$\tau = \tau (n, {\bf t}, \bar {\bf t})$
which satisfies an infinite set of bilinear differential-difference
equations of Hirota type \cite{DJKM83,JM83}. 

It turns out that it is possible to impose some
constraints on the Lax operators in such a way that they would be
consistent with the dynamics of the Toda hierarchy. One of such
examples was considered in our previous work \cite{KZ21a}, where we have
introduced the Toda hierarchy with a constraint of type C, which is
$\bar L=L^{\dag}$ (in the symmetric gauge). Here and below $L^{\dag}$
is the conjugate operator 
($(f(n)\circ e^{\p_n})^{\dag}=e^{-\p_n}\circ f(n)$). 
It is a subhierarchy of the Toda lattice which can be regarded as
an integrable discretization of the CKP hierarchy \cite{DJKM81}-\cite{KZ21}. 

The purpose of 
this paper is to elaborate another example and to
introduce a new integrable hierarchy: the Toda hierarchy with the
constraint of type B.
In a special
gauge, which we call the balanced gauge, it reads:
\beq\label{int1}
L^{\dag}=(e^{\p_n}-e^{-\p_n})\bar L (e^{\p_n}-e^{-\p_n})^{-1}.
\eeq
This constraint is preserved by the flows $\p_{t_k}-\p_{\bar t_k}$ and is
destroyed by the flows $\p_{t_k}+\p_{\bar t_k}$, so to define the 
hierarchy one should restrict the times as $\bar t_k=-t_k$. 
This hierarchy is an integrable discretization of the 
BKP hierarchy \cite{DJKM81,DM-H09,DJKM82,DJKM82a,LW99,Tu07,Z21} which
is defined by imposing the constraint
\beq\label{int2}
(L^{\rm KP})^{\dag}=-\p_x L^{\rm KP} \p_x^{-1}, \qquad x=t_1
\eeq
on the pseudo-differential Lax operator of the KP hierarchy, that is why
we call (\ref{int1}), looking as a difference analogue of
(\ref{int2}), ``the constraint of type B''. 
The first member of the hierarchy is the following system of equations
for two unknown functions $v$, $f_0$:
\beq\label{int3}
\left \{\begin{array}{l}
\displaystyle{ \p_{t_1}\log \Bigl (v(n)v(n+1)\Bigr )=
\frac{f_0(n+1)}{v(n+1)}}-\frac{f_0(n)}{v(n)},
\\ \\
\p_{t_2}v(n)-\p_{t_1}f_0(n)=2v^2(n)\Bigl (v(n-1)-v(n+1)\Bigr ).
\end{array} \right.
\eeq
Let us note that essentially the same hierarchy was suggested 
in the paper \cite{GK10} as an integrable discretization of the 
Novikov-Veselov equation. However, the close connection with the
Toda lattice was not mentioned there. 

We construct algebraic-geometrical (quasi-periodic) solutions of the 
Toda lattice with the constraint of type B. These solutions are built from 
algebraic curves with holomorphic involution $\iota$ having exactly two
fixed points $Q_1, Q_2$ (ramified coverings with two branch points) and
two marked points $P_0$, $P_{\infty}$ such that $\iota (P_0)=P_{\infty}$.
We show that the solutions can be expressed in terms of Prym theta-functions.
A similar but different construction was given in \cite{GK10}, where
unramified coverings with two pairs of marked points were considered.

Solutions to the Toda lattice with the constraint of type B can be 
expressed in terms of the tau-function $\tau^B(n)=
\tau^B(n, {\bf t})$ as follows:
\beq\label{int4}
v(n)=\frac{\tau^B(n+1)\tau^B(n-1)}{(\tau^B(n))^2},
\quad
f_0(n)=v(n)\, \p_{t_1}\! \log \frac{\tau^B(n+1)}{\tau^B(n-1)}.
\eeq
The tau-function $\tau^B(n)$ of the Toda lattice with the constraint
is related to the tau-function $\tau (n)$ of the Toda lattice as
\beq\label{int4a}
\tau (n)=\tau^B(n) \tau^B(n-1).
\eeq
This relation should be compared with the relation between tau-functions
of the KP and BKP hierarchies: the former is the square of the latter.
In the Toda case, it is not the square but the arguments of the two 
factors are shifted by $1$; in the continuum limit they become the same.

We show that
for algebraic-geometrical solutions of equations (\ref{int3}), 
constructed starting from a smooth algebraic curve $\Gamma$ with
holomorphic involution (having exactly two fixed points) 
and a divisor satisfying
some special condition (see (\ref{al4}) below),
the tau-function $\tau^B(n)$ 
is expressed in terms of Prym theta-function 
$\Theta_{\rm Pr}$:
\beq\label{int5}
\tau^B(n, {\bf t})=e^{-L(n, {\bf t})-\frac{1}{2}\, Q(n, {\bf t})}\,
\Theta_{\rm Pr}\Bigl (n\vec u_0 +\sum_{k\geq 1} \vec u_k t_k +\vec z\Bigr ),
\eeq
where $L(n, {\bf t})$, $Q(n, {\bf t})$ are linear and quadratic forms
in $n$, ${\bf t}$ respectively and $\vec u_i$ are periods of certain 
differentials on the algebraic curve $\Gamma$.  

To avoid a confusion, we should stress that our Toda hierarchies
with constraints of types C and B introduced in \cite{KZ21a} and in this
paper respectively, are very different from what is called C- and 
B-Toda in \cite{UT84}. 

In section 2, after a short reminder about the general Toda lattice,
we introduce the constraint of type B and prove that it is consistent with
the hierarchy if one restricts the time variables 
to the submanifold $t_k+\bar t_k=0$
in the space of independent variables. The first nontrivial equations
of the hierarchy are obtained in the explicit form. Section 3
is devoted to the construction of algebraic-geometrical solutions in terms
of Prym theta-functions. The main technical tool is the Baker-Akhiezer
function. All necessary facts about algebraic curves, differentials and
theta-functions are given in the appendix. 

\section{The Toda hierarchy of type B}

\subsection{2D Toda lattice}

First of all, we briefly review the 2D Toda lattice hierarchy following \cite{UT84}.
Let us consider the pseudo-difference Lax operators
\beq\label{toda1}
{\cal L}=e^{\p_n}+\sum_{k\geq 0}U_k(n) e^{-k\p_n}, \quad
\bar {\cal L}=c(n)e^{-\p_n}+\sum_{k\geq 0}\bar U_k(n) e^{k \p_n},
\eeq
where $e^{\p_n}$ is the shift operator acting as
$e^{\pm \p_n}f(x)=f(n\pm 1)$. 
The coefficient functions $c(n)$, $U_k(n)$, $\bar U_k(n)$
are functions of $n$ and all the times ${\bf t}$, $\bar {\bf t}$.
The Lax equations are
\beq\label{toda2}
\begin{array}{l}
\p_{t_m}{\cal L}=[{\cal B}_m +\Phi_m, {\cal L}], \quad
\p_{t_m}\bar {\cal L}=[{\cal B}_m +\Phi_m, \bar {\cal L}],
\quad {\cal B}_m=({\cal L}^m)_{>0}, \quad \Phi_m =({\cal L}^m)_{0},
\\ \\
\p_{\bar t_m}{\cal L}=[\bar {\cal B}_m, {\cal L}], \quad
\p_{\bar t_m}\bar {\cal L}=[\bar {\cal B}_m, \bar {\cal L}],
\quad \bar {\cal B}_m=(\bar {\cal L}^m)_{< 0}.
\end{array}
\eeq
Here and below, given a subset $\SSS \subset \ZZ$, we denote
$\displaystyle{\Bigl (\sum_{k\in \z} U_k e^{k\eta \p_x}\Bigr )_{\sss}=
\sum_{k\in \sss} U_k e^{k\eta \p_x}}$.
For example, 
\beq\label{toda4a}
\begin{array}{l}
{\cal B}_1=e^{\p_n}, \qquad
\bar {\cal B}_1=c(n)e^{-\p_n},
\\ \\
{\cal B}_2=e^{2\p_n}+(U_0(n)+U_0(n+1))e^{\p_n},
\\ \\
\bar {\cal B}_2=c(n)c(n-1)e^{-2\p_n}+c(n)(\bar U_0(n)+
\bar U_0(n-1))e^{-\p_n},
\\ \\
\Phi_1=U_0(n), \qquad \Phi_2=U_1(n)+U_1(n+1)+U_0^2(n).
\end{array}
\eeq
We will also use the notation 
$(\ldots )_{+}=(\ldots )_{>0}$, $(\ldots )_{-}=(\ldots )_{<0}$.
Setting
\beq\label{toda4}
c(n)=e^{\varphi (n)-\varphi (n-1)},
\eeq
we have from (\ref{toda2}):
\beq\label{toda3}
\p_{t_m}\varphi =\Phi_m, \quad
\p_{\bar t_m}\varphi =-(\bar {\cal L}^m)_0 :=-\bar \Phi_m.
\eeq
In terms of the function $\varphi (n)$ the first equation of the Toda
lattice hierarchy has the form
\beq\label{toda5}
\p_{t_1}\p_{\bar t_1}\varphi (n)=e^{\varphi (n)-\varphi (n-1 )}-
e^{\varphi (n+1)-\varphi (n)}.
\eeq
The common solution of the hierarchy is given by the tau-function
$\tau (n) = \tau (n, {\bf t}, \bar {\bf t})$. In particular, 
\beq\label{toda5a}
e^{\varphi (n)}=\frac{\tau (n+1)}{\tau (n)}.
\eeq

There exist other equivalent formulations of the Toda hierarchy
obtained from the one given above by gauge 
transformations \cite{Takebe1,Takebe2}.
So far we have used the standard gauge in which
the coefficient of the first term of ${\cal L}$ is fixed to be $1$. 
In fact
there is a family of gauge transformations with a function $g=g(n)$
of the form
$$
{\cal L}\to g^{-1}{\cal L}g , \quad \bar {\cal L}\to g^{-1}\bar {\cal L}g,
$$
$$
{\cal B}_n \to g^{-1}{\cal B}_n g -g^{-1}\p_{t_n}g, \quad
\bar {\cal B}_n \to g^{-1}\bar {\cal B}_n g -g^{-1}\p_{\bar t_n}g.
$$
We are interested in another special gauge in which the coefficients
in front of the first terms of the two Lax operators coincide.
Let us denote the Lax operators and the generators of the flows
in this gauge by $L$, $\bar L$, $B_m$, $\bar B_m$ respectively:
\beq\label{toda6}
\begin{array}{c}
L=g^{-1}{\cal L}g, \quad \bar L=g^{-1}\bar {\cal L}g,
\\ \\
B_m =g^{-1}{\cal B}_m g +\Phi_m -\p_{t_m}\log g , \quad
\bar B_m =g^{-1}\bar {\cal B}_m g -\p_{\bar t_m}\log g.
\end{array}
\eeq
It is easy to see that the function $g(n)$ is determined from the relation
\beq\label{toda7}
e^{\varphi (n)}=g(n)g(n+1).
\eeq
We call this gauge the balanced gauge.

An equivalent formulation of the Toda hierarchy is through the 
Zakharov-Shabat equations
\beq\label{toda8}
\begin{array}{l}
\p_{t_n}B_m -\p_{t_m}B_n +[B_m, B_n]=0,
\\ \\
\p_{\bar t_n}B_m -\p_{t_m}\bar B_n +[B_m, \bar B_n]=0,
\\ \\
\p_{\bar t_n}\bar B_m -\p_{\bar t_m}\bar B_n +[\bar B_m, \bar B_n]=0.
\end{array}
\eeq
They are compatibility conditions for the auxiliary linear problems
\beq\label{toda9}
\p_{t_m}\Psi =B_m\Psi , \quad \p_{\bar t_m}\Psi =\bar B_m\Psi .
\eeq
The wave function $\Psi =\Psi (n, {\bf t}, \bar {\bf t}, k)$ depends
on a spectral parameter $k\in \CC$. We will often 
skip the dependence on ${\bf t}$, $\bar {\bf t}$ writing 
simply $\Psi =\Psi (n,k)$. The wave function has the following expansions
as $k\to \infty$ and $k\to 0$:
\beq\label{toda10}
\Psi (n, k) = \left \{
\begin{array}{l}
\displaystyle{
k^n e^{\xi ({\bf t}, k)}e^{\varphi_{+}(n)}\Bigl (1+
\sum_{s\geq 1} \xi_s(n)k^{-s}\Bigr ), \quad k\to \infty ,}
\\ \\
\displaystyle{
k^n e^{\xi (\bar {\bf t}, k^{-1})}e^{\varphi_{-}(n)}\Bigl (1+
\sum_{s\geq 1} \chi_s(n)k^{s}\Bigr ), \quad k\to 0 ,}
\end{array}
\right.
\eeq
where $\displaystyle{\xi ({\bf t}, k)=\sum_{j\geq 1}t_j k^j}$.

It is convenient to represent the wave function as a result of
acting of the dressing operators $W$, $\bar{W}$ to the exponential
function
$$
E(n,{\bf t},k)=k^n e^{\xi ({\bf t}, k)}.
$$
The dressing operators are pseudo-difference operators of the form
\beq\label{toda11}
\begin{array}{l}
\displaystyle{
W=e^{\varphi_{+}(n)}\Bigl (1+
\sum_{s\geq 1} \xi_s(n)e^{-s\p_n }\Bigr ),}
\\ \\
\displaystyle{
\bar{W}=e^{\varphi_{-}(n)}\Bigl (1+
\sum_{s\geq 1} \chi_s(n)e^{s\p_n }\Bigr ),}
\end{array}
\eeq
so that
\beq\label{toda12}
\Psi (n, k) = \left \{
\begin{array}{l}
\displaystyle{WE(n, {\bf t},k), \quad k\to \infty} ,
\\ \\
\displaystyle{\bar{W}E(-n, \bar {\bf t},k^{-1}), \quad k\to 0.}
\end{array}
\right.
\eeq
The Lax operators are obtained by ``dressing'' of the shift operators
as follows:
\beq\label{toda13}
L=We^{\p_n}W^{-1}, \quad \bar L=\bar W e^{-\p_n}\bar W^{-1}.
\eeq
It is clear from (\ref{toda13}) that the wave function is an eigenfunction
of the operators $L$, $\bar L$:
\beq\label{toda14}
L\Psi (n,k)  =k\Psi (n,k), \quad \bar L \Psi (n,k)=k^{-1}\Psi (n,k).
\eeq

Let us introduce the dual wave function $\Psi^*$ as
\beq\label{toda15a}
\Psi^{*}(n,k)=(W^{\dag})^{-1}E^{-1}(n, {\bf t}, k), \quad k\to \infty ,
\eeq
where conjugation of difference operators (the $\dag$-operation) is defined
on shift 
operators as $$(f(n)\circ e^{\p_n})^{\dag}=e^{-\p_n}\circ f(n)$$
and is extended to all difference and pseudo-difference operators 
by linearity. The following lemmas are well-known.

\begin{lemma}
The dual wave function satisfies the conjugate linear equations
\beq\label{toda16a}
\begin{array}{c}
L^{\dag}\Psi^* (n,k)=k\Psi^{*}(n,k), 
\\ \\
-\p_{t_m}\Psi^*(n,k)=B_m^{\dag}\Psi^*(n,k), 
\quad
-\p_{\bar t_m}\Psi^*(n,k)=\bar B_m^{\dag}\Psi^*(n,k).
\end{array}
\eeq
\end{lemma}

\begin{lemma}
The wave function and its dual satisfy the bilinear relation
\beq\label{toda17}
\oint_{C_{\infty}}\Psi (n, {\bf t}, \bar {\bf t}, k)
\Psi^* (n', {\bf t}', \bar {\bf t}', k)\frac{dk}{k}=
\oint_{C_{0}}\Psi (n, {\bf t}, \bar {\bf t}, k)
\Psi^* (n', {\bf t}', \bar {\bf t}', k)\frac{dk}{k}
\eeq
for all $n, n'$, ${\bf t}, \, {\bf t}'$, $\bar {\bf t}, \, \bar {\bf t}'$,
where $C_{\infty}$, $C_0$ are small contours around $\infty$, $0$ 
respectively.
\end{lemma}

\subsection{The constraint of type B}

Let us define the operator
\beq\label{c1}
T=e^{-\varphi (n)}e^{\p_n}
\eeq
and consider the following constraint on the two Lax operators
in the standard gauge:
\beq\label{c2}
(T-T^{\dag})\bar {\cal L}={\cal L}^{\dag}(T-T^{\dag}).
\eeq
As is easy to see, in the balanced gauge this constraint acquires the form
\beq\label{c3}
(e^{\p_n}-e^{-\p_n})\bar L =L^{\dag}(e^{\p_n}-e^{-\p_n}).
\eeq

\begin{theorem}\label{Theorem:constraint}
The constraint (\ref{c2}) is invariant under the 
flows $\p_{t_k}\! -\! \p_{\bar t_k}$
for all $k\geq 1$.
\end{theorem}

\noindent
{\it Proof.} We should prove that
$$
(\p_{t_k}\! -\! \p_{\bar t_k})\Bigl [
(T-T^{\dag})\bar {\cal L}-{\cal L}^{\dag}(T-T^{\dag})\Bigr ]=0.
$$
Basically, this is a straightforward calculation which uses the Lax
equations and the equations $\p_{t_k}T=-\Phi_k T$, 
$\p_{\bar t_k}T=\bar \Phi_k T$. Here are some details.
Denoting
$$
{\cal A}_k={\cal B}_k -\bar {\cal B}_k,
$$
we can write, after some cancellations:
\beq\label{c3a}
\begin{array}{c}
(\p_{t_k}\! -\! \p_{\bar t_k})\Bigl [
(T-T^{\dag})\bar {\cal L}-{\cal L}^{\dag}(T-T^{\dag})\Bigr ]
\\ \\
=\Bigl ((T\! -\! T^{\dag}){\cal A}_k +{\cal A}_k^{\dag}(T\! -\! T^{\dag})
\Bigr )\bar {\cal L}-{\cal L}^{\dag}
\Bigl ((T\! -\! T^{\dag}){\cal A}_k +{\cal A}_k^{\dag}(T\! -\! T^{\dag})
\Bigr )
\\ \\
-\Phi_k T\bar {\cal L}-\bar \Phi_k T \bar {\cal L}
+T^{\dag}\bar \Phi_k\bar {\cal L}
+{\cal L}^{\dag}\bar \Phi_k T -{\cal L}^{\dag}T^{\dag}\bar \Phi_k
-{\cal L}^{\dag}T^{\dag}\Phi_k
+T\Phi_k \bar {\cal L} -T \bar {\cal L}\Phi_k
\\ \\
+T^{\dag}\bar {\cal L}\Phi_k +\Phi_k {\cal L}^{\dag}T 
-\Phi_k {\cal L}^{\dag}T^{\dag}+{\cal L}^{\dag}\Phi_k T^{\dag}.
\end{array}
\eeq
Now, writing
$$
(T-T^{\dag})\bar {\cal L}^k-({\cal L}^{\dag})^k(T-T^{\dag})=0
$$
and taking the $(\ldots )_{-}$-part of this equality,
we get
\beq\label{c4}
(T-T^{\dag})\bar {\cal B}_k-{\cal B }_k^{\dag}(T-T^{\dag})=
- (({\cal L}^{\dag})^k)_{-1}e^{-\p_n}T
+T(\bar {\cal L}^k)_{-1}e^{-\p_n}
-\Phi_k T^{\dag}+T^{\dag}\bar \Phi_k.
\eeq
Similarly, writing
$$
(T-T^{\dag}){\cal L}^k-(\bar {\cal L}^{\dag})^k(T-T^{\dag})=0
$$
and taking the $(\ldots )_{+}$-part of this equality,
we get
\beq\label{c4a}
(T-T^{\dag}){\cal B}_k-\bar {\cal B }_k^{\dag}(T-T^{\dag})=
((\bar {\cal L}^{\dag})^k)_{1}e^{\p_n}T^{\dag} 
-T^{\dag}({\cal L}^k)_{1}e^{\p_n}
+\bar \Phi_k T-T\Phi_k.
\eeq
Subtracting (\ref{c4}) from (\ref{c4a}) and taking into account that
$$
T(\bar {\cal L}^k)_{-1}e^{-\p_n}=
((\bar {\cal L}^{\dag})^k)_{1}e^{\p_n}T^{\dag}, \quad
(({\cal L}^{\dag})^k)_{-1}e^{-\p_n}T=
T^{\dag}({\cal L}^k)_{1}e^{\p_n},
$$
we obtain:
$$
(T\! -\! T^{\dag}){\cal A}_k +{\cal A}_k^{\dag}(T\! -\! T^{\dag})
=\bar \Phi_k T-T \Phi_k +\Phi_k T^{\dag}-T^{\dag}\bar \Phi_k.
$$
Plugging this into (\ref{c3a}) we see that all terms in the right hand side
cancel and we get zero.
\square

We have proved that the constraint (\ref{c2}) 
remains intact under the flows $\p_{t_k}\! -\! \p_{\bar t_k}$.
However, it is destroyed by the flows  $\p_{t_k}\! +\! \p_{\bar t_k}$.

The invariance of the constraint proved in the standard gauge implies that
the constraint in the balanced gauge is invariant, too, i.e.
\beq\label{c6}
(\p_{t_k}-\p_{\bar t_k})\Bigl [(e^{\p_n}-e^{-\p_n})\bar L-L^{\dag}
(e^{\p_n}-e^{-\p_n})\Bigr ]=0.
\eeq
In the balanced 
gauge, the generators of the flows $\p_{t_k}-\p_{\bar t_k}$ are
\beq\label{c5}
A_k=B_k-\bar B_k.
\eeq
A straightforward calculation which uses the Lax equations allows one
to prove that (\ref{c6}) implies
\beq\label{c7}
(e^{\p_n}-e^{-\p_n})A_k+A_k^{\dag}(e^{\p_n}-e^{-\p_n})=0
\eeq
or
\beq\label{c8}
A_k^{\dag}=-(e^{\p_n}-e^{-\p_n})A_k(e^{\p_n}-e^{-\p_n})^{-1}.
\eeq
As soon as $A_k^{\dag}$ is a difference operator (a linear combination
of a finite number of shifts), this relation suggests that the operator
$A_k$ must be divisible by $e^{\p_n}-e^{-\p_n}$ from the right:
\beq\label{c9}
A_k =C_k (e^{\p_n}-e^{-\p_n}),
\eeq
where $C_k$ is a difference operator. The substitution into (\ref{c7})
then shows that $C_k$ is a self-adjoint operator: $C_k=C_k^{\dag}$. 

\subsection{Equations of the Toda lattice of type B}

Let us introduce the following linear combinations of times:
\beq\label{toda15}
T_j = \frac{1}{2} (t_j-\bar t_j), \quad y_j = \frac{1}{2} (t_j+\bar t_j),
\eeq
then the corresponding vector fields are
\beq\label{toda16}
\p_{T_j}=\p_{t_j}-\p_{\bar t_j}, \quad
\p_{y_j}=\p_{t_j}+\p_{\bar t_j}.
\eeq
We have seen that the $T_j$-flows preserve the constraint while the
$y_j$-flows destroy it. This suggests to put $y_j=0$ for all $j$ and 
consider the evolution with respect to $T_j$ only. In this way one can
introduce the Toda hierarchy of type B. The balanced gauge
is most convenient for this purpose. 
According to (\ref{c9}), the generators of the $T_k$-flows
in this gauge have the general form
\beq\label{b1}
A_k=\Bigl (f_{0k} +\sum_{j=1}^{k-1}(f_{jk}e^{j\p_n}+e^{-j\p_n}
f_{jk})\Bigr )(e^{\p_n}-e^{-\p_n}),
\eeq
where $f_{jk}$ are some functions of $n$ and the times $\{T_i\}$. 
The equations of the hierarchy are obtained from the Zakharov-Shabat
conditions
$$
[\p_{T_k}-A_k, \, \p_{T_m}-A_m]=0
$$
or
\beq\label{b2}
\p_{T_m}A_k-\p_{T_k}A_m +[A_k, A_m]=0.
\eeq

The simplest equations are obtained when $k=1$, $m=2$. In this case
the operators are
\beq\label{b3}
\begin{array}{l}
A_1= v(n)(e^{\p_n}-e^{-\p_n}),
\\ \\
A_2= \Bigl (f_0(n) +f_1(n) e^{\p_n}+e^{-\p_n}f_1(n)\Bigr )
(e^{\p_n}-e^{-\p_n}).
\end{array}
\eeq
Plugging these into the Zakharov-Shabat equation
$$
\p_{T_2}A_1-\p_{T_1}A_2 +[A_1, A_2]=0,
$$
we obtain the following system of three equations for three 
unknown functions:
\beq\label{b4}
\left \{\begin{array}{l}
v(n)f_1(n+1)=v(n+2)f_1(n),
\\ \\
\p_{T_1}f_1(n)+v(n+1) f_0(n)-v(n)f_0(n+1)=0,
\\ \\
\p_{T_2}v(n)-\p_{T_1}f_0(n)+2v (n)\Bigl (f_1(n)-f_1(n-1)\Bigr )=0.
\end{array}
\right.
\eeq
The first equation allows one to exclude the function $f_1$:
$$
f_1(n)=v(n)v(n+1),
$$
and the system becomes
\beq\label{b5}
\left \{\begin{array}{l}
\displaystyle{ \p_{T_1}\log \Bigl (v(n)v(n+1)\Bigr )=
\frac{f_0(n+1)}{v(n+1)}-
\frac{f_0(n)}{v(n)}},
\\ \\
\p_{T_2}v(n)-\p_{T_1}f_0(n)+2v^2(n)\Bigl (v(n+1)-v(n-1)\Bigr )=0.
\end{array} \right.
\eeq

Let us compare the balanced gauge with the standard gauge explicitly. 
Using equations (\ref{toda4a}) and (\ref{toda6}), we can write:
\beq\label{b6}
A_1=\frac{g(n+1)}{g(n)}\, e^{\p_n} +U_0(n)-\p_{T_1}\log g(n)-
\frac{g(n-1)}{g(n)}\, e^{\varphi (n)-\varphi (n-1)}e^{-\p_n},
\eeq
\beq\label{b6a}
\begin{array}{c}
\displaystyle{
A_2=\frac{g(n+2)}{g(n)}\, e^{2\p_n} +
\frac{g(n+1)}{g(n)}(U_0(n)+U_0(n+1))e^{\p_n}}
\\ \\
\displaystyle{ +
U_1(n)+U_1(n+1) +U_0^2(n)-\p_{T_2}\log g(n)}
\\ \\
\displaystyle{
-\frac{g(n-1)}{g(n)}\, e^{\varphi (n)-\varphi (n-1)}
(U_0(n)+U_0(n+1))e^{-\p_n}-
\frac{g(n-2)}{g(n)}\, e^{\varphi (n)-\varphi (n-2)}
e^{-2\p_n}},
\end{array}
\eeq
where we have put $\bar U_0(n)=U_0(n+1)$, as it follows from the 
constraint. Comparing with (\ref{b3}), we identify
\beq\label{b7}
v(n)=\frac{g(n+1)}{g(n)},
\eeq
\beq\label{b8}
f_0(n)=\frac{g(n+1)}{g(n)}\, (U_0(n)+U_0(n+1))
\eeq
and
\beq\label{b9}
\p_{T_1}\log g(n)=U_0(n).
\eeq

The tau-function of the Toda lattice with the constraint, 
$\tau^B(n)=\tau^B(n, {\bf T})$, can be introduced via the relation
\beq\label{b10}
g(n)=\frac{\tau^B(n)}{\tau^B(n-1)},
\eeq
then
\beq\label{b11}
v(n)=\frac{\tau^B(n+1)\tau^B(n-1)}{(\tau^B(n))^2}, \quad
f_0(n)=v(n)\, \p_{T_1}\! \log \frac{\tau (n+1)}{\tau (n-1)}.
\eeq
After these substitutions, the first of equations (\ref{b5}) 
is satisfied identically while the second one reads
\beq\label{b11a}
\begin{array}{l}
\displaystyle{
\p_{T_2}\! \log \frac{\tau^B(n-1)\tau^B(n+1)}{(\tau^B(n))^2}
+\p_{T_1}^2 \! \log \frac{\tau^B(n-1)}{\tau^B(n+1)}}
\\ \\
\phantom{aaaaaaaaaaa}\displaystyle{
+\, \p_{T_1}\! \log \frac{\tau^B(n-1)\tau^B(n+1)}{(\tau^B(n))^2}\,
\p_{T_1}\! \log \frac{\tau^B(n-1)}{\tau^B(n+1)}}
\\ \\
\phantom{aaaaaaaaaaaaaaaaaaaa}\displaystyle{
=\, 2\frac{\tau^B(n-2)\tau^B(n+1)}{\tau^B(n-1)\tau^B(n)}-2
\frac{\tau^B(n+2)\tau^B(n-1)}{\tau^B(n+1)\tau^B(n)}}.
\end{array}
\eeq
Note that this equation is cubic in $\tau^B$. 

As it follows from equations (\ref{toda5a}), (\ref{toda7}), 
the tau-function $\tau^B(n)$ is connected with
the tau-function of the Toda lattice $\tau (n)$ by the relation
\beq\label{b12}
\tau (n, {\bf t}, -{\bf t})=\tau^B(n-1, {\bf t})\, \tau^B(n, {\bf t}).
\eeq
This is the Toda analogue of the well known relation between the KP and
BKP tau-functions: the former is the square of the latter. 

\begin{figure}[t]
\centering{\includegraphics[scale=1.2]{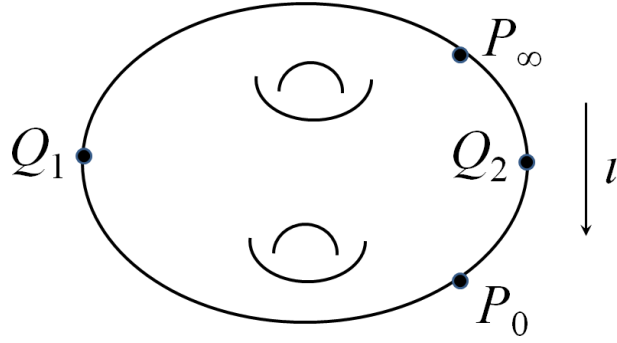}}
\vspace{1cm}
\caption{Algebraic curve $\Gamma$ with involution.}
\label{figure:curve}
\end{figure}

\section{Algebraic-geometrical solutions}

Consider algebraic-geometrical solutions of the Toda hierarchy 
constructed from a genus $g$ algebraic curve admitting 
holomorphic involution $\iota$ with two fixed
points $Q_1$, $Q_2$ and an effective degree $g$ divisor
satisfying the condition (\ref{al4}) below. 
Let $P_{\infty}$, $P_0$ be two marked 
points (different from $Q_1$, $Q_2$) such that $P_0=\iota P_{\infty}$
with local parameters $k^{-1}$ and $k=\iota (k^{-1})$ respectively
(Fig. \ref{figure:curve}). 
In this section we denote
the set of independent variables of the hierarchy as ${\bf t}=
(t_1, t_2, \ldots )$ (we set $\bar t_k =-t_k$ in the full 2D Toda 
hierarchy and put $T_k=t_k$). Set
\beq\label{al1}
U^{\alpha}({\bf t})=\sum_{j\geq 1} t_j(U_j^{\alpha}+U_j^{\alpha +g_0})
\quad \mbox{or} \quad 
\vec U({\bf t})=\sum_{j\geq 1}t_j (\vec U_j-\iota \vec U_j).
\eeq
Here and below we use the notation introduced in the appendix.

The Baker-Akhiezer function has the form
\beq\label{al2}
\begin{array}{l}
\displaystyle{
\Psi (n,{\bf t}, P)=\exp \Bigl (\sum_k t_k(\Omega_k(P)-\Omega_k (\iota P))+
n \Omega_0(P)\Bigr )}
\\ \\
\phantom{aaaaaaaaaaaaaaaaaaa}\displaystyle{\times \, 
\frac{\Theta (\vec A(P)+n\vec U_0 +\vec U({\bf t})+\vec Z)
\Theta (\vec Z)}{\Theta (\vec A(P)+\vec Z)
\Theta (n\vec U_0 +\vec U({\bf t})+\vec Z)}},
\end{array}
\eeq
where the vector $\vec Z$ is given by
\beq\label{al3}
\vec Z =-\vec A({\cal D})-\vec K.
\eeq
Here ${\cal D}$ is an effective non-special divisor of degree $g$ and
$\vec K$ is the vector of Riemann's constants. 
The function $\Psi (n, {\bf t}, P)$ has poles at the points of the
divisor ${\cal D}$. 
We assume that the divisor ${\cal D}$ is subject to the condition
\beq\label{al4}
{\cal D}+\iota {\cal D}={\cal K}+Q_1+Q_2,
\eeq
where ${\cal K}$ is the canonical class. This relation is equivalent to
\beq\label{aaa}
\vec A({\cal D})+\vec A(\iota {\cal D})+2\vec K=\vec A(Q_1)+\vec A(Q_2).
\eeq
Note that with out choice of the initial point of the Abel map we have
$\vec A(Q_1)=0$.
The Baker-Akhiezer function (\ref{al2}) is normalized in such a way that
\beq\label{al5}
\Psi (n,{\bf t}, Q_1)=1.
\eeq
Near the points $P_{\infty}$, $P_0$ it has essential singularities 
of the form
\beq\label{al6}
\begin{array}{l}
\Psi (n,{\bf t}, P)=e^{\varphi_+(n)}k^n e^{\xi ({\bf t}, k)}\Bigl (1+
O(k^{-1})\Bigr ), \quad P\to P_{\infty} \quad (k\to \infty ),
\\ \\
\Psi (n,{\bf t}, P)=e^{\varphi_-(n)}k^n e^{-\xi ({\bf t}, k^{-1})}\Bigl (1+
O(k)\Bigr ), \quad P\to P_{0} \quad (k\to 0 ).
\end{array}
\eeq 
From the explicit formula (\ref{al2}) we have
\beq\label{al7}
\begin{array}{l}
\displaystyle{
e^{\varphi_+(n)}=\frac{\Theta (\vec A(P_{\infty})+n\vec U_0 +
\vec U({\bf t})+\vec Z)
\Theta (\vec Z)}{\Theta (\vec A(P_{\infty})+\vec Z)
\Theta (n\vec U_0 +\vec U({\bf t})+\vec Z)}\,
\exp \Bigl (-\! \sum_k \Omega_k(P_0)t_k\Bigr )},
\\ \\
\displaystyle{
e^{\varphi_-(n)}=\frac{\Theta (\vec A(P_{0})+n\vec U_0 +
\vec U({\bf t})+\vec Z)
\Theta (\vec Z)}{\Theta (\vec A(P_{0})+\vec Z)
\Theta (n\vec U_0 +\vec U({\bf t})+\vec Z)}\,
\exp \Bigl (\, \sum_k \Omega_k(P_0)t_k\Bigr )}.
\end{array}
\eeq

The dual Baker-Akhiezer function $\Psi^*(n, {\bf t}, P)$ can be introduced
as follows. Let $d\Omega (P)$ be the meromorphic differential with simple poles
at the points $P_0$, $P_{\infty}$ with residues 
$\pm 1$ and zeros at the points of the divisor ${\cal D}$. This differential
has other $g$ zeros at the points of some effective divisor ${\cal D}^*$
such that
\beq\label{a14b}
{\cal D}+{\cal D}^*={\cal K}+P_0+P_{\infty},
\eeq
or
\beq\label{al4c}
\vec A({\cal D})+\vec A({\cal D}^*)+2\vec K=\vec A(P_0)+\vec A(P_{\infty}).
\eeq
The dual Baker-Akhiezer function is a unique function with poles at
the points of the divisor ${\cal D}^*$ and essential singularities
of the form
\beq\label{al6a}
\begin{array}{l}
\Psi^* (n,{\bf t}, P)=e^{-\varphi_+(n)}k^{-n} e^{-\xi ({\bf t}, k)}\Bigl (1+
O(k^{-1})\Bigr ), \quad P\to P_{\infty} \quad (k\to \infty ),
\\ \\
\Psi^* (n,{\bf t}, P)=e^{-\varphi_-(n)}k^{-n} e^{\xi ({\bf t}, k^{-1})}
\Bigl (1+
O(k)\Bigr ), \quad P\to P_{0} \quad (k\to 0 ).
\end{array}
\eeq 

\begin{theorem}
The Baker-Akhiezer function and its dual obey the bilinear relation
\beq\label{b1a}
\res_{P=P_0}\Bigl (\Psi (n, {\bf t}, P)\Psi^* (n', {\bf t}', P)
d\Omega (P)\Bigr )+
\res_{P=P_{\infty}}\Bigl (\Psi (n, {\bf t}, P)
\Psi^* (n', {\bf t}', P)d\Omega (P)\Bigr )=0
\eeq
for all $n, n'$, ${\bf t}, {\bf t'}$.
\end{theorem}

\noindent
{\it Proof.} The differential $\Psi (n, {\bf t}, P)\Psi^* (n', {\bf t}', P)
d\Omega (P)$ is a well-defined differential on $\Gamma$. 
From the definition of $d\Omega$ it follows that it has singularities
only at the points $P_0$, $P_{\infty}$. Therefore, sum of the residues
must be zero.
\square

\begin{remark}
The bilinear relation (\ref{toda17}) is the realization of (\ref{b1a})
written in the local parameter.
\end{remark}

Let $d\hat \Omega (P)$ be the meromorphic differential with simple poles
at the points $Q_1$, $Q_2$ with residues $\pm 1$ and zeros at the points
of the divisor ${\cal D}+\iota {\cal D}$. Such differential is unique.
It is given by the explicit formula
\beq\label{al8}
\begin{array}{l}
\displaystyle{
d\hat \Omega (P)=\frac{C \, \Theta_*(\vec A(Q_2)-\vec A(Q_1))}{\Theta 
(\vec Z)\Theta (
\vec A(Q_1)-\vec A(Q_2)-\vec Z)}}
\\ \\
\phantom{aaaaaaaaaaaaaaa}\displaystyle{
\times \, \frac{\Theta (\vec A(P)-\vec A(Q_1)+\vec Z)
\Theta (\vec A(P)-\vec A(Q_2)-\vec Z)}{\Theta_*(\vec A(P)-\vec A(Q_1))
\Theta_*(\vec A(P)-\vec A(Q_2))}}\,
\, d\zeta (P) ,
\end{array}
\eeq
where $C$ is a constant depending only on the point $P_{\infty}$ and
$d\zeta$ is the holomorphic differential
\beq\label{al9}
d\zeta (P)=\sum_{\alpha =1}^g \Theta_{* , \alpha }(0)d\omega_{\alpha}(P)
\eeq
and $\Theta_{*}$ is the odd theta-function with some odd half-integer
characteristics. It is known \cite{Mumford} 
that the differential $d\zeta$ has 
$g-1$ double zeros at the points, where the function 
$\Theta_*(\vec A(P)-\vec A(Q))$ has simple 
zeros other than the zero at $P=Q$
(they are independent of $Q$).

Consider the differential
$$
d\hat \Omega (n, {\bf t}, P)=\Psi (n, {\bf t}, P)\Psi (n, {\bf t}, \iota P)
d\hat \Omega (P).
$$
It is a well-defined meromorphic differential with simple poles
at the points $Q_1$, $Q_2$ and no other singularities. Therefore,
\beq\label{al10}
1=\res_{P=Q_1}d\hat 
\Omega (n, {\bf t}, P)=-\res_{P=Q_2}d\hat \Omega (n, {\bf t}, P)=
\Psi^2 (n, {\bf t}, Q_2),
\eeq
i.e., $\Psi (n, {\bf t}, Q_2)=\pm 1$. 

\begin{lemma}
It holds
\beq\label{al11}
\Psi (n, {\bf t}, Q_2)=(-1)^n.
\eeq
\end{lemma}

\noindent
{\it Proof.} First consider the case $n=0$. Since the Baker-Akhiezer
function 
is continuous in ${\bf t}$ and $\Psi (0, 0, Q_2)=1$, we conclude that
$\Psi (0, {\bf t}, Q_2)=1$. Next consider the case $n=1$.
The correct sign can be determined by the following
argument. The function $\Psi (1, {\bf t}, Q_2)$ 
is continuous in ${\bf t}$
and $P_{\infty}$ for $P_{\infty}\neq Q_1$. 
Therefore, in order to find the correct sign,
we set ${\bf t}=0$ and choose $P_{\infty}$ in a neighborhood of 
$Q_1$ with some local parameter $z^{-1}$
which is odd under the involution. By the definition of $\Omega_0(P)$ we have
$$
e^{\Omega_0(P)}=\frac{z(P)-z(P_{\infty})}{z(P)+z(P_{\infty})}\, 
\Bigl (1+ O(z^{-1}(P))\Bigr ).
$$
For any fixed positive $r\ll 1$ one can choose $\eta$ such that for 
$|z^{-1}(P_{\infty})|<\eta$
and $|z^{-1}(P)|=r$ the inequality
$
\Bigl |1+e^{\Omega_0(P)}\Bigr |<M\eta
$
holds with a constant $M$ independent of $\eta$. 
Moreover, for any fixed path from a point
$P'$, $|z^{-1}(P')|=r$, to $Q_2$ there exists a constant 
$M_1$ (which depends on the path
but does not depend on $\eta$) such that the inequality
$\displaystyle{\Bigl |\int_{P'}^{Q_2}d\Omega_0 \Bigr |<M_1\eta}$ holds. It then follows
that for a fixed path from $Q_1$ to $Q_2$
\beq\label{pr4}
\lim_{P_{\infty}\to Q_1}e^{\Omega_0(Q_2)}=-1.
\eeq
In fact if we define $\displaystyle{\tilde \Omega_0(P)=
\int_{Q_0}^Pd\Omega_0}$ with
some initial point $Q_0\to Q_1$, then the result depends on the order of
the limits:
$$
\lim_{Q_0\to Q_1}\lim_{P_{\infty}\to Q_1}e^{\tilde \Omega_0(Q_2)}=1
\quad
\mbox{but}
\quad
\lim_{P_{\infty}\to Q_1} \lim_{Q_0\to Q_1}e^{\tilde \Omega_0(Q_2)}=-1,
$$
and our case is the latter limit.
The ratio of theta-functions is a continuous function of 
$P_{\infty}$ which
tends to $1$ as $P_{\infty}\to Q_1$ and the formula (\ref{al11})
for $n=1$ is proved. 
The statement of the lemma for arbitrary $n$ follows from equation
(\ref{pr4}).
\square

The following theorem is the key for constructing the algebraic-geometrical
solutions of the Toda hierarchy with the constraint of type B.

\begin{theorem}\label{theorem:identity}
Let $\Gamma$ be an algebraic curve with involution $\iota$ having
two fixed points $Q_1, Q_2$, and let $\vec Z$ be a vector such that
\beq\label{b2a}
\vec Z +\iota \vec Z =-\vec A(Q_1)-\vec A(Q_2).
\eeq 
Then the following identity holds:
\beq\label{b3a}
\begin{array}{l}
\Theta^2(\vec A(P)
\! +\! \vec Z)
=c(P)\Theta (\vec A(P)\! -\! \vec A(\iota P)
\! +\! \vec Z)\Theta (\vec Z),
\end{array}
\eeq
where $c(P)$ is some constant depending only on $P$.
\end{theorem}

\noindent
{\it Proof.}
Consider the differential
$$
d\tilde \Omega (n, {\bf t}, P)=
\Psi (n+1, {\bf t}, P)\Psi (n, {\bf t}, \iota P)
d\Omega (P).
$$
It is a well-defined meromorphic differential with simple poles
at the points $Q_1$, $Q_2$, $P_{\infty}$ and 
no other singularities. The sum of its residues must be zero. 
Therefore,
\beq\label{al12a}
\res_{P=P_{\infty}}d\tilde \Omega (n, {\bf t}, P)=2
\eeq
since the residues at the points $Q_{1,2}$ are equal to $-1$.
We have:
\beq\label{al12}
\res_{P=P_{\infty}}d\tilde \Omega (n, {\bf t}, P)=
e^{\varphi_{+}(n+1)+\varphi_{-}(n)}\, 
\frac{\tilde C \, \Theta_*(\vec A(Q_2))
\Theta (\vec A(P_{\infty})\! +\! \vec Z)
\Theta (\vec A(P_{\infty}) \! -\! 
\vec A(Q_2)\! -\! \vec Z)}{\Theta_*(\vec A(P_{\infty}))
\Theta_*(\vec A(P_{\infty})-\vec A(Q_2))
\Theta (\vec Z)\Theta (\vec A(Q_2)\! +\! \vec Z)}
\eeq
with some constant $\tilde C$ depending only on $P_{\infty}$.
Substituting here $e^{\varphi_{\pm}(n)}$ from (\ref{al7})
and using (\ref{qp11}), we obtain:
\beq\label{al13}
\frac{\tilde C(P_{\infty})\,
A(P_{\infty}, \vec Z)\, \Theta^2(\vec A(P_0)+n \vec U_0 +\vec U({\bf t})
+\vec Z)}{\Theta (\vec A(P_0)-\vec A(\iota P_0)+n \vec U_0 +\vec U({\bf t})
+\vec Z)\Theta (n \vec U_0 +\vec U({\bf t})
+\vec Z)}=2,
\eeq
where
\beq\label{al14}
A(P_{\infty}, \vec Z)=\frac{\Theta (\vec A(P_{\infty})-\vec A(Q_2)-
\vec Z)\Theta (\vec Z)}{\Theta (\vec A(Q_2)+
\vec Z)\, \Theta (\vec A(P_{0})+\vec Z)}.
\eeq
Some additional arguments based on the 
behavior of the theta-functions under involution 
show that in fact $A(P_{\infty}, \vec Z)=1$. Therefore, we have obtained
the identity
\beq\label{al15}
\begin{array}{l}
\Theta^2(\vec A(P_0)\! +\! n \vec U_0 \! +\! \vec U({\bf t})
\! +\! \vec Z)
\\ \\
\phantom{aaaaaaaaaaaa}=c(P_0)\Theta (\vec A(P_0)\! -\! 
\vec A(\iota P_0)\! +\! 
n \vec U_0 \! +\! \vec U({\bf t})
\! +\! \vec Z)\Theta (n \vec U_0 \! +\! \vec U({\bf t})
\! +\! \vec Z).
\end{array}
\eeq
Now, putting $P_0=P$, $n={\bf t}=0$ and noting that (\ref{aaa}) 
with $\vec Z =\vec A({\cal D})-\vec K$ implies that 
$\vec Z +\iota \vec Z =-\vec A(Q_1)-\vec A(Q_2)$, we arrive at 
(\ref{b3a}). Note that $\iota \vec U_0=-\vec U_0$, 
$\iota \vec U({\bf t})=-\vec U({\bf t})$, so the vector
$\vec Z_{n, {\bf t}}=\vec Z + n\vec U_0+\vec U({\bf t})$ 
still satisfies the condition (\ref{b2a}).
\square

\begin{corollary}
Under the assumptions of Theorem \ref{theorem:identity}, it holds:
\beq\label{al16}
\Theta (\vec A(P)
\! +\! \vec Z)=c_1(P)\, \Theta_{\rm Pr}(\vec A^{\rm Pr}(P)+\vec z)
\, \Theta_{\rm Pr}(\vec z),
\eeq
where $\vec z=\sigma^{-1}(\vec Z +\frac{1}{2}\vec A(Q_2))$
and the map $\sigma$ is defined in (\ref{ind1}).
\end{corollary}

\noindent
{\it Proof.} The condition (\ref{aaa}) can be rewritten as
$$
\begin{array}{c}
\iota \Bigl (\vec Z +\frac{1}{2}\vec A(Q_2)\Bigr )=-
\Bigl (\vec Z +\frac{1}{2}\vec A(Q_2)\Bigr ),
\end{array}
$$
so the map 
$\sigma^{-1}(\vec Z +\frac{1}{2}\vec A(Q_2)):=\vec z\in \CC^{g_0}$ is
well-defined. The same is true for the map 
$\sigma^{-1}(\vec A(P)-\vec A(\iota P)+
\vec Z +\frac{1}{2}\vec A(Q_2))=\vec A^{\rm Pr}(P)+\vec z$, where
$\vec A^{\rm Pr}(P)$ is the Abel-Prym map (\ref{inv3a}). Then,
taking the square root of both sides of identity (\ref{b3a}) and using
(\ref{inv8}), we arrive at (\ref{al16}).
\square

Restoring the dependence on $n$ and ${\bf t}$ in (\ref{al16}),
we can write it in the form
\beq\label{al17}
\Theta (\vec A(P_0)\! +\! n \vec U_0 \! +\! \vec U({\bf t})
\! +\! \vec Z)=c_1(P_0)\, \Theta_{\rm Pr}((n+1)\vec u_0\! +\! 
\vec u({\bf t})\! +\! 
\vec z)\, \Theta_{\rm Pr}(n\vec u_0\! +\! \vec u({\bf t})\! +\! 
\vec z),
\eeq
where $\vec u_0=\vec A^{\rm Pr}(P_0)$, $\vec u({\bf t})=\sigma^{-1}
(\vec U({\bf t}))$.

\begin{theorem}\label{theorem:bilinear}
The Baker-Akhiezer function (\ref{al2}) satisfies the bilinear identity
\beq\label{b101}
\begin{array}{l}
\displaystyle{
\res_{P=P_0}\Bigl (\Psi (n, {\bf t}, P)\Psi (n', {\bf t}', \iota P)
d\hat \Omega (P)\Bigr )}
\\ \\
\phantom{aaaaaaaaaaaaaaaaa}\displaystyle{+
\res_{P=P_{\infty}}\Bigl (\Psi (n, {\bf t}, P)
\Psi (n', {\bf t}', \iota P)d\hat \Omega (P)\Bigr )=1-(-1)^{n-n'}}
\end{array}
\eeq
for all $n, n'$, ${\bf t}, {\bf t'}$.
\end{theorem}

\noindent
{\it Proof.} The differential
$\Psi (n, {\bf t}, P)
\Psi (n', {\bf t}', \iota P)d\hat \Omega (P)$ is a well-defined 
differential on $\Gamma$ with essential singularities at the points 
$P_0$, $P_{\infty}$ and simple poles at the points $Q_1$, $Q_2$.
Equation (\ref{b101}) is the statement that sum of its residues 
is equal to zero.
\square

The next theorem establishes the connection between the dual 
Baker-Akhiezer function $\Psi^*(n,P)$ and the function $\Psi (n, \iota P)$.

\begin{theorem}\label{theorem:psi}
The following identity holds:
\beq\label{b102}
\Psi^*(n,P)d\Omega (P)=A\Bigl (\Psi (n\! +\! 1, \iota P)-
\Psi (n\! -\! 1, \iota P)\Bigr )d\hat \Omega (P)
\eeq
with some constant $A$.
\end{theorem}

\begin{remark}
Equation (\ref{b102}) 
implies that subtracting the bilinear identities of the form 
(\ref{b101}) taken at
$n'+1$ and $n'-1$, we get the bilinear identity (\ref{b1a}). 
\end{remark}

\noindent
{\it Proof of Theorem \ref{theorem:psi}.} 
Let us show that poles and zeros of the differentials
in both sides of (\ref{b102}) coincide. Both sides 
of (\ref{b102}) have essential singularities at the points $P_0, P_{\infty}$
and the exponential factors at these singularities coincide.
Besides, there is a pole of order $n+1$ at $P_0$
in the left hand side of (\ref{b102}), and the differential is holomorphic
in all other points. It is easy to see that the differential in the 
right hand side has the same singularity. 
Possible poles at $Q_1$, $Q_2$ cancel because
$\Psi (n\! +\! 1, Q_{1,2})=
\Psi (n\! -\! 1, Q_{1,2})$ (see (\ref{al11})).
As far as zeros are concerned,
both sides have a zero of order $n-1$ (here we assume that $n\geq 1$)
and $g$ simple zeros at the points of the divisor ${\cal D}$. 
The linear space of such differentials is one-dimensional, whence 
the two sides of (\ref{b102}) are proportional to each other.
In order to find the coefficient of proportionality, we compare the
leading terms of $\Psi^*(n,P)$ and 
$\Psi (n\! +\! 1, \iota P)-
\Psi (n\! -\! 1, \iota P)$ at $P_0$. The coefficients
at the leading terms are $e^{-\varphi_{-}(n)}$ and
$e^{\varphi_{+}(n+1)}$ respectively. 
From the proof of Theorem \ref{theorem:identity} 
(see (\ref{al12a}), (\ref{al12})) it follows that
\beq\label{b105}
e^{\varphi_{+} (n+1)+\varphi_{-}(n)}=\mbox{const}.
\eeq
Therefore, the coefficient of proportionality $A$ 
does not depend on $n$. 
\square

The Baker-Akhiezer function is an eigenfunction of the Lax operators:
\beq\label{b103}
L\Psi (n,P)=k\Psi (n,P), \qquad \bar L\Psi (n,\iota P)=k\Psi (n,\iota P)
\eeq
and
\beq\label{b104}
L^{\dag}\Psi^*(n, P)=k \Psi^*(n, P).
\eeq
It is easy to see that our normalization of the Baker-Akhiezer function
corresponds to the balanced gauge. Indeed, writing $L=a(n)e^{\p_n}+\ldots$,
$\bar L=\bar a(n)e^{-\p_n} +\ldots$, and comparing the leading terms
in (\ref{b103}), we have:
$$
a(n)e^{\varphi_+(n+1)-\varphi_+(n)}=1, \quad
\bar a(n)e^{\varphi_-(n-1)-\varphi_-(n)}=1.
$$
Therefore, 
$$
\frac{\bar a(n)}{a(n)}=e^{\varphi_-(n)-\varphi_-(n-1)-\varphi_+(n)
+\varphi_+(n+1)}=1
$$
due to (\ref{b105}). The following corollary from Theorem \ref{theorem:psi}
states that the $L$-operators obey the constraint of type B.

\begin{corollary}
The constraint (\ref{c3}) for the $L$-operators of the Toda lattice 
in the balanced gauge holds.
\end{corollary}

\noindent
{\it Proof.} 
Using (\ref{b102}), we can write
$$
\Psi^*(n,P)=A'(P)\Bigl (\Psi (n\! +\! 1, \iota P)-
\Psi (n\! -\! 1, \iota P)\Bigr ),
$$
where
$\displaystyle{A'(P)=A\frac{d\hat \Omega (P)}{d\Omega (P)}}$. Therefore
(see (\ref{b103}), (\ref{b104})),
$$
\bar L(n+1)\Psi (n+1, \iota P)-\bar L(n-1)\Psi (n-1, \iota P)
=k\Bigl (\Psi (n+1, \iota P)-\Psi (n-1, \iota P)\Bigr )
$$
$$
=k(A'(P))^{-1}\Psi^*(n, P)=(A'(P))^{-1}L^{\dag}(n)\Psi^*(n, P)
$$
$$
=L^{\dag}(n)\Bigl (\Psi (n+1, \iota P)-\Psi (n-1, \iota P)\Bigr ).
$$
or
$$
\Bigl (\bar L(n+1)-L^{\dag}(n)\Bigr )\Psi (n+1, \iota P)=
\Bigl (\bar L(n-1)-L^{\dag}(n)\Bigr )\Psi (n-1, \iota P).
$$
Since this is true for any $P$, the equality should hold for the 
operators, i.e.,
$$
\Bigl (\bar L(n+1)-L^{\dag}(n)\Bigr )e^{\p_n}=
\Bigl (\bar L(n-1)-L^{\dag}(n)\Bigr )e^{-\p_n},
$$
which is (\ref{c3}).
\square

Comparing with the standard gauge, we can write:
\beq\label{b106}
e^{\varphi (n)}=e^{\varphi_-(n)-\varphi_+(n)}=g(n)g(n+1).
\eeq
In terms of the tau-function we have:
\beq\label{b107}
e^{\varphi (n)}=\frac{\tau (n+1)}{\tau (n)}=\frac{\tau^B(n+1)}{\tau^B(n-1)},
\eeq
where $\tau^B(n)=\tau^B(n, {\bf t})$ 
is the tau-function of the Toda hierarchy
with the constraint of type B introduced in (\ref{b10}). 
This tau-function is connected with
the tau-function of the Toda lattice by the relation (\ref{b12}).

The algebraic-geometrical solutions of the Toda lattice were 
constructed by one of the authors 
in \cite{Krichever81} using the general method 
suggested in \cite{Krichever77,Krichever77a}. Restricting to the 
subspace $t_k +\bar t_k=0$, we can write the tau-function:
\beq\label{b108}
\tau (n, {\bf t})=e^{-L(n, {\bf t})-Q(n, {\bf t})}\Theta \Bigl (
\vec A(P_{\infty})+n\vec U_0 +\vec U({\bf t})+\vec Z\Bigr ),
\eeq
where $L(n, {\bf t})$ is an inessential linear form in $n$, ${\bf t}$ and
$Q(n, {\bf t})$ is the quadratic form
\beq\label{b109}
Q(n, {\bf t})=\sum_{i,j}(\Omega_{ij}-\omega_{ij})t_it_j -2n\sum_j
\Omega_j(P_0)t_j
\eeq
($\Omega_{ij}$ and $\omega_{ij}$ are coefficients of expansions
of the function $\Omega_i(P)$ near $P_{\infty}$ and $P_0$, see
(\ref{qp3b}), (\ref{qp6c})). Then $e^{\varphi (n)}$
is given by
\beq\label{b110}
e^{\varphi (n)}=\exp \Bigl (2\sum_j \Omega_j(P_0)t_j\Bigr )
\frac{\Theta (\vec A(P_{\infty})+\vec Z)}{\Theta (\vec A(P_{0})+\vec Z)}\,
\frac{\Theta (\vec A(P_0)+n\vec U_0 +\vec U({\bf t})
+\vec Z)}{\Theta (\vec A(P_{\infty})+n\vec U_0 +\vec U({\bf t})
+\vec Z)}.
\eeq
Using equation (\ref{al17}), we find the tau-function of the Toda lattice
with the constraint:
\beq\label{b111}
\tau^B(n, {\bf t})=e^{-L^B(n, {\bf t})-\frac{1}{2}\, Q(n, {\bf t})}
\Theta_{\rm Pr}\Bigl (n\vec u_0 +\vec u({\bf t})+\vec z\Bigr ),
\eeq
where $\vec u_0=\vec A^{\rm Pr}(P_0)$, $\vec u({\bf t})=\sigma^{-1}
(\vec U({\bf t}))$ and $L^B(n, {\bf t})$ is an inessential
linear form. The formulas
\beq\label{b112}
v(n)=\frac{\tau^B(n+1)\tau^B(n-1)}{(\tau^B(n))^2},
\eeq
\beq\label{b113}
f_0(n)=\frac{\tau^B(n+1)\tau^B(n-1)}{(\tau^B(n))^2}\, \p_{t_1}\!
\log \frac{\tau^B(n+1)}{\tau^B(n-1)}
\eeq
provide a solution to equations
(\ref{b5}).

\begin{remark}
For algebraic-geometrical solutions, the discrete variable $n$ 
can be regarded
as a continuous variable, as equation (\ref{b111}) 
suggests. The Baker-Akhiezer
functions then have a discontinuity on a cut between the 
points $P_0$ and $P_{\infty}$. The right hand side 
of equation (\ref{al11}) should then be $e^{i\pi n}$.  
\end{remark}

\section{Concluding remarks}

We have introduced a subhierarchy of the 2D Toda lattice by imposing
the constraint of the form (\ref{c3}) on the two Lax operators. 
Restricting the dynamics to the subspace $t_k+\bar t_k=0$ of the 
space of independent variables, we have shown that this constraint
is invariant under flows of the Toda hierarchy. The hierarchy which 
is obtained in this way can be regarded as an integrable discretization
of the BKP hierarchy. We have also constructed its algebraic-geometrical
solutions in terms of Prym theta-functions. 

Along with the CKP and BKP hierarchies, in \cite{DJKM81} a whole
family of subhierarchies of KP indexed by $m\in \ZZ_{\geq 0}$
was introduced 
of which $m=0$ ($m=1$) case corresponds to the BKP (CKP) hierarchy.
They are defined by imposing a constraint of the type 
\beq\label{conc1}
(L^{\rm KP})^{\dag}=-Q^{-1}_m L^{\rm KP} Q_m
\eeq
on the Lax operator $L^{\rm KP}$ of the KP hierarchy. Here
$$Q_m=\p_x^{m-1}+\mbox{lower order terms}$$
is a difference operator of order $m-1$ 
such that
\beq\label{conc2}
Q_m^{\dag}=(-1)^{m-1}Q_m
\eeq
(at $m=0$ one sets $Q_0=\p_x^{-1}$). It was also shown that in the case
$m=2$ ($Q_2=\p_x$) one obtains a hierarchy which is equivalent to BKP.
To the best of our knowledge, nothing is known about the cases $m\geq 3$.

It would be interesting to investigate whether more general constraints 
analogous to (\ref{conc1}) exist in the Toda case. Hypothetically, 
they might have the form
\beq\label{}
L^{\dag}=S^{-1}_m \bar L S_m,
\eeq
where $S_m$ is a difference operator (a finite linear 
combination of powers of $e^{\p_n}-e^{-\p_n}$ with some 
coefficients) of order $m-1$ such that
$$
S_m^{\dag}=(-1)^{m-1}S_m.
$$
At $m=0$ we set
$S_0=(e^{\p_n}-e^{-\p_n})^{-1}$, then the case $m=0$ ($m=1$) corresponds
to the Toda hierarchy with the constraint of type B discussed in this paper
(respectively, the Toda hierarchy with the constraint of type C \cite{KZ21a}).
Presumably, in the case $m=2$ one obtains a hierarchy which is 
equivalent to the Toda lattice of type B. 

\section*{Appendix: Algebraic curves, differentials
and theta-functions}
\addcontentsline{toc}{section}{Appendix: Algebraic curves, differentials
and theta-functions}
\def\theequation{A\arabic{equation}}
\def\theHequation{\theequation}
\setcounter{equation}{0}

Here we present the basic notions and facts related to  
algebraic curves (Riemann surfaces) \cite{Springer,Fay}
and theta-functions \cite{Mumford}
which are necessary for the construction of quasi-periodic solutions to the 
Toda hierarchy.

\paragraph{Matrix of periods.}
Let $\Gamma$ be a smooth compact algebraic curve  
of genus $g$. We fix a canonical basis of
cycles ${\sf a}_{\alpha}, {\sf b}_{\alpha}$ ($\alpha =1, \ldots , g$) with the intersections
${\sf a}_{\alpha}\circ {\sf a}_{\beta}={\sf b}_{\alpha}\circ {\sf b}_{\beta}=0$,
${\sf a}_{\alpha}\circ {\sf b}_{\beta}=\delta_{\alpha \beta}$ and a basis of holomorphic 
differentials $d\omega_{\alpha}$  
normalized by the condition
$\displaystyle{\oint_{{\sf a}_{\alpha}}d\omega_{\beta}=\delta_{\alpha \beta}}$. 
The matrix of periods is defined as
\beq\label{qp1}
T_{\alpha \beta}=\oint_{{\sf b}_{\alpha}}d\omega_{\beta}, \qquad \alpha , \beta =1, \ldots , g.
\eeq
It is a symmetric matrix with positively defined imaginary part.
The Jacobian of the curve $\Gamma$ is the $g$-dimensional complex torus
\beq\label{qp4}
J(\Gamma )=\CC ^g /\{\vec N +T \vec M\},
\eeq
where $\vec N$, $\vec M$ are $g$-dimensional vectors with integer components.

\paragraph{Riemann theta-functions.}
The Riemann theta-function associated with the Riemann surface is defined by the 
absolutely convergent series
\beq\label{qp2}
\Theta(\vec z)=\Theta(\vec z|T)=
\sum_{\vec n \in \z ^{g}}e^{\pi i (\vec n, T\vec n)+2\pi i (\vec n, \vec z)},
\eeq
where $T$ is the matrix of periods, 
$\vec z=(z_1, \ldots , z_g)$ and $\displaystyle{(\vec n, \vec z)=
\sum_{\alpha =1}^g n_{\alpha}z_{\alpha}}$.
It is an entire function with the following quasi-periodicity property:
\beq\label{qp2a}
\Theta (\vec z +\vec N +T\vec M )=\exp (-\pi i (\vec M, T\vec M)-
2\pi i (\vec M, \vec z)) \Theta (\vec z).
\eeq
More generally, one can introduce the theta-functions with characteristics
$\vec \delta'$, $\vec \delta''$:
\beq\label{char}
\Theta \left [\begin{array}{c}\vec \delta ' \\ \vec \delta^{''}
\end{array}\right ] (\vec z)
=\sum_{\vec n \in \z ^{g}}e^{\pi i (\vec n +\vec \delta ', 
T(\vec n +\vec \delta '))+2\pi i (\vec n+\vec \delta ', 
\vec z+\vec \delta '')}.
\eeq
Half-integer characteristics play an especially important role.
Let $\displaystyle{\left [
\begin{array}{c}\vec \delta ' \\ \vec \delta ''
\end{array}\right ]}$ be an odd half-integer 
characteristics, i.e. such that $(-1)^{4((\vec \delta ', \vec \delta '')}=-1$.
For brevity, by $\Theta_{*}$ we denote the corresponding 
odd theta-function:
$\displaystyle{\Theta \left [\begin{array}{c}\vec \delta ' \\ \vec \delta^{''}
\end{array}\right ] (\vec z)}:=\Theta_{*}(\vec z)$, 
$\Theta_{*}(-\vec z)=-\Theta_{*}(\vec z)$.

\paragraph{The Abel map.}
The Abel map $\vec A(P)$, $P\in \Gamma$
from $\Gamma$ to $J(\Gamma )$ is defined as
\beq\label{qp3}
\vec A(P)=
\int_{Q_1}^P d \vec \omega , \qquad d\vec \omega =(d\omega_1, 
\ldots , d\omega_g ).
\eeq
The Abel map can be extended to the group of divisors ${\cal D}=n_1P_1+\ldots +n_KP_K$ as
\beq\label{qp3a}
\vec A({\cal D})=\sum_{i=1}^K n_i\int_{Q_1}^{P_i} d \vec \omega =\sum_{i=1}^K
n_i\vec A(P_i).
\eeq

Consider the function $f(P)=\Theta (\vec A(P)-\vec e)$ and assume that it is not 
identically zero. It can be shown that this function has $g$ zeros on $\Gamma$
at a divisor ${\cal D}=Q_1+ \ldots +Q_g$ and $\vec A( {\cal D})=\vec e -\vec K$,
where $\vec K=
(K_1, \ldots , K_g)$ is the 
vector of Riemann's constants
\beq\label{qp7}
K_{\alpha}= \pi i +\pi i T_{\alpha \alpha}-2\pi i
\sum_{\beta \neq \alpha}\oint_{a_{\beta}}\omega_{\alpha} (P)d\omega_{\beta}(P).
\eeq
In other words,
for any non-special effective divisor
${\cal D}=Q_1+\ldots +Q_g$ of degree $g$ the function
$$
f(P)=\Theta \Bigl (\vec A(P)-\vec A({\cal D}) -\vec K\Bigr )
$$
has exactly $g$ zeros at the points $Q_1, \ldots , Q_g$.  
Let ${\cal K}$ be the canonical class of divisors (the equivalence class of divisors
of poles and zeros of abelian differentials on $\Gamma$), then one can show that
\beq\label{qp8}
2\vec K=-\vec A({\cal K}).
\eeq
It is known that $\mbox{deg}\, {\cal K}=2g-2$. In particular, this means that 
holomorphic differentials have $2g-2$ zeros on $\Gamma$. 

\paragraph{The Riemann bilinear identity.}
Let ${\sf a}_{\alpha}$, ${\sf b}_{\alpha}$ be Jordan arcs 
which represent the canonical basis of cycles
intersecting transversally at a single point and let $$\tilde 
\Gamma =\Gamma \setminus \bigcup_{\alpha}
\Bigl ({\sf a}_{\alpha}\bigcup {\sf b}_{\alpha}\Bigr )$$ be a simply 
connected subdomain of $\Gamma$
(a fundamental domain). Integrals of any two 
meromorphic differentials $d\Omega '$,
$d\Omega ''$ satisfy the Riemann bilinear identity
\beq\label{r1}
\oint_{\p \tilde \Gamma}\Bigl (\int_{Q_1}^P d\Omega ' \Bigr )d\Omega ''(P)=
\sum_{\alpha =1}^g \left ( \oint_{{\sf a}_{\alpha}}d\Omega '\oint_{{\sf b}_{\alpha}}d\Omega ''-
\oint_{{\sf b}_{\alpha}}d\Omega '\oint_{{\sf a}_{\alpha}}d\Omega ''\right ).
\eeq
In particular, if ${\sf a}$-periods of both differentials are equal to zero, we have
\beq\label{r2}
\oint_{\p \tilde \Gamma}\Bigl (\int_{Q_1}^P 
d\Omega ' \Bigr )d\Omega ''(P)=0.
\eeq

\paragraph{Differentials of the second and third kind.}
Let $P_{\infty}\in \Gamma$ be a marked point and $k^{-1}$ a local parameter 
in a neighborhood of the marked point ($k=\infty$ at $P_{\infty}$). 
Let $d\Omega_j$ 
be differentials of the second kind with the only pole 
at $P_{\infty}$ of the form
$$
d\Omega_j = dk^j +O(k^{-2})dk, \quad k\to \infty
$$
normalized by the condition 
$\displaystyle{\oint_{{\sf a}_{\alpha}}d\Omega_{j}=0}$, and $\Omega_j$
be the (multi-valued) functions
$$
\Omega_j(P)=\int_{Q_1}^{P}d\Omega_j +q_j,
$$
where the constants $q_j$ are chosen in such a way that 
$\Omega_i(P)=k^i +O(k^{-1})$, namely,
\beq\label{qp3b}
\Omega_i(P)=k^i +\sum_{j\geq 1} \frac{1}{j}\, \Omega_{ij}k^{-j}.
\eeq
It follows from the Riemann bilinear identity 
that the matrix $\Omega_{ij}$ is symmetric:
$\Omega_{ij}=\Omega_{ji}$ 
(one should put $d\Omega '=d\Omega_i$, $d\Omega ''=d\Omega_j$
in (\ref{r2})).

Set
\beq\label{qp5}
U_j^{\alpha}=\frac{1}{2\pi i}\oint_{{\sf b}_{\alpha}}d\Omega_j, \qquad
\vec U_j =(U_j^{1}, \ldots , U_j^g).
\eeq
One can prove the following relation:
\beq\label{qp6a}
d\vec \omega =\sum_{j\geq 1}\vec U_j k^{-j-1}dk
\eeq
or
\beq\label{qp6}
\vec A(P)-\vec A(P_{\infty})=\int_{P_{\infty}}^P d\vec \omega =
-\sum_{j\geq 1}\frac{1}{j}\, \vec U_j k^{-j}
\eeq
(this follows from (\ref{r1}) if one puts $d\Omega '=d\omega_{\alpha}$,
$d\Omega ''=d\Omega_j$). 

Similarly, 
let $P_{0}\in \Gamma$ be another marked point with local parameter $k$ ($k(P_0)=0$).
Let $d\bar \Omega_j$ 
be differentials of the second kind with the only pole at $P_{0}$ of the form
$$
d\bar \Omega_j = dk^{-j} +O(1)dk, \quad k\to 0
$$
normalized by the condition 
$\displaystyle{\oint_{{\sf a}_{\alpha}}d\bar \Omega_{j}=0}$, 
and 
$$
\bar \Omega_i(P)=\int_{Q_0}^{P}d\bar \Omega_i +\bar q_i=
k^{-i} +\sum_{j\geq 1} \frac{1}{j}\, \bar \Omega_{ij}k^{j}, 
\quad \bar \Omega_{ij}=\bar \Omega_{ji}.
$$
Set
\beq\label{qp5a}
\bar U_j^{\alpha}=\frac{1}{2\pi i}\oint_{{\sf b}_{\alpha}}d\bar \Omega_j.
\eeq

We will also need 
expansions of the functions $\Omega_i(P)$, $\bar \Omega_i(P)$ near
the points $P_0$, $P_{\infty}$ respectively:
\beq\label{qp6c}
\begin{array}{l}
\displaystyle{
\Omega_i(P)=\Omega_i (P_0)+\sum_{j\geq 1}\frac{1}{j}\, \omega_{ij}k^j, \quad P\to P_0,}
\\ \\
\displaystyle{
\bar \Omega_i(P)=\bar \Omega_i (P_\infty )+
\sum_{j\geq 1}\frac{1}{j}\, \bar \omega_{ij}k^{-j}, \quad P\to P_\infty ,}
\end{array}
\eeq
and the Riemann bilienar identity implies that 
$\bar \omega_{ij}=\omega_{ji}$.

Let $d\Omega_0$ be the meromorphic dipole 
differential (of the third kind) 
with zero ${\sf a}$-periods having a simple pole at 
$P_0$ with residue $+1$ and 
another simple pole at $P_{\infty}$ with residue $-1$. 
Set
\beq\label{qp10}
U_0^{\alpha}=\frac{1}{2\pi i}\oint_{{\sf b}_{\alpha}}d\Omega_0.
\eeq
The following important relation is an immediate consequence of the Riemann bilinear
identity:
\beq\label{qp11}
\vec A(P_0)-\vec A(P_{\infty})=U_0.
\eeq

\paragraph{Curves with involution.}
Assume now that the curve $\Gamma$ 
admits a holomorphic involution $\iota$ with two fixed
points $Q_1$ and $Q_2$. 
The Riemann-Hurwitz formula implies that genus $g$ is even:
$g=2g_0$. The curve $\Gamma$ is two-sheet covering of the factor-curve
$\Gamma_0=\Gamma /\iota$ of genus $g_0$. 
The canonical basis of cycles 
on $\Gamma$ can be chosen in such a way that
$\iota {\sf a}_{\alpha}=-{\sf a}_{g_0+\alpha}$, 
$\iota {\sf b}_{\alpha}=-{\sf b}_{g_0+\alpha}$, 
$\alpha =1, \ldots , g_0$. Here and below
the sums like $\alpha +g_0$ for $\alpha =1, \ldots , g$ 
are understood modulo $g$, i.e., for example,
$(g_0+1)+g_0=1$.
Then the normalized holomorphic differentials 
obey the properties $\iota^{*}d\omega_{\alpha}=
-d\omega_{g_0+\alpha}$, $\alpha =1, \ldots , g_0$. With this choice, the matrix
of periods enjoys the symmetry
\beq\label{inv0}
T_{\alpha \beta}=T_{\alpha +g_0, \beta +g_0}.
\eeq
The involution $\iota$ induces the involution of the 
Jacobian: for $\vec u =(u_1, \ldots , u_g)$ we have $\iota u_{\alpha}
=-u_{\alpha +g_0}$, or, explicitly,
\beq\label{ind}
\iota \vec u=
\iota (u_1, \ldots , u_g)=-(u_{g_0+1}, \ldots , u_g, u_1, \ldots , u_{g_0}).
\eeq
Note that the symmetry property (\ref{inv0}) of the period matrix implies the relation
\beq\label{inv0a}
\Theta (\iota \vec u)=\Theta (\vec u) \quad \mbox{for any $\vec u\in \CC^g$},
\eeq
where $\Theta$ is the Riemann theta-function. 

\paragraph{The Prym variety.}
The Prym variety $Pr(\Gamma )\subset J(\Gamma )$ is a subvariety of the Jacobian
of $\Gamma$ defined by the condition $\iota (\vec z)=-\vec z$, i.e. it is the 
variety
$$
Pr(\Gamma )=
\Bigl \{ \vec z \in \CC^g \Bigr | \, \vec z =(z_1, \ldots , z_{g_0}, 
z_1, \ldots , z_{g_0})
\Bigr \}/(\ZZ^g + T\ZZ^g).
$$

The Prym differentials are
defined as $d\upsilon_{\alpha}=d\omega_{\alpha}+d\omega_{g_0+\alpha}$
(here $\alpha =1, \ldots , g_0$); they are odd
with respect to the involution. The $g_0\times g_0$ matrix
\beq\label{inv1}
\Pi_{\alpha \beta}=\oint_{{\sf b}_{\alpha}}d\upsilon_{\beta}, \quad \alpha =1, \ldots , g_0
\eeq
is called the Prym matrix of periods. It is a symmetric matrix with positively 
defined imaginary part. 

The Prym variety is isomorphic to the $g_0$-dimensional torus
$
Pr=\CC^{g_0}/((\ZZ^{g_0} + \Pi \ZZ^{g_0}).
$
This torus can be embedded into the Jacobian by the map
\beq\label{ind1}
\sigma (\vec u)=(u_1, \ldots , u_{g_0}, u_1, \ldots , u_{g_0})\in J(\Gamma ),
\quad \vec u =(u_1, \ldots , u_{g_0}),
\eeq
and the image is the Prym variety. 

\paragraph{The Abel-Prym map.}
The Abel-Prym map $\vec A^{\rm Pr}(P)$, $P\in \Gamma$
from $\Gamma$ to $Pr(\Gamma )$ is defined as
\beq\label{inv3}
\vec A^{\rm Pr}(P)=
\int_{Q_1}^P d \vec \upsilon , \qquad d\vec \upsilon =
(d\upsilon_1, \ldots , d\upsilon_{g_0}),
\eeq
or
\beq\label{inv3a}
A^{{\rm Pr}}_{\alpha}(P)=A_{\alpha}(P)+A_{g_0+\alpha}(P)=
A_{\alpha}(P)-A_{\alpha}(\iota P).
\eeq
The Abel-Prym map can be extended to the group of divisors by linearity.
Note that for curves with involution the initial point of the Abel map is 
not arbitrary: it is $Q_1$, one of the two fixed points of the involution. 
Note that according to our definitions 
$A_{\alpha}(\iota P)=-A_{g_0+\alpha}(P)$ that
agrees with the induced involution of the Jacobian (\ref{ind}). 

\paragraph{The Prym theta-functions.}
The Prym theta-function is defined by the series
\beq\label{inv2}
\Theta_{\rm Pr}(\vec z)=\Theta_{\rm Pr}(\vec z|\Pi )=
\sum_{\vec n \in \z ^{g_0}}e^{\pi i (\vec n, \Pi \vec n)+2\pi i 
(\vec n, \vec z)}.
\eeq
There is a remarkable relation between the 
Riemann and Prym theta-functions \cite{Fay}:
for any $\vec u\in \CC^{g_0}$
it holds
\beq\label{inv8}
\begin{array}{c}
\Theta \Bigl (\sigma (\vec u)+\frac{1}{2}\vec A(Q_2)\Bigr )=
\Theta \Bigl (\sigma (\vec u)-\frac{1}{2}\vec A(Q_2)\Bigr )
=C \Bigl (\Theta_{\rm Pr}(\vec u)\Bigr )^2,
\end{array}
\eeq
i.e., the Riemann theta-function on this part of the Jacobian is a full square
and the square root of it 
is, up to a constant multiplier $C$, the Prym theta-function.

\section*{Acknowledgments}

\addcontentsline{toc}{section}{Acknowledgments}

The research of A.Z. has been funded within the framework of the
HSE University Basic Research Program.

\end{document}